\documentclass[]{aa}

\usepackage{txfonts}
\usepackage{graphicx}
\usepackage{amsmath}
\usepackage{xcolor}
\usepackage{placeins}
\usepackage{ragged2e}

\begin{document}

\title{Sunyaev-Zel'dovich detection of the galaxy cluster Cl J1449+0856 at 
$z=1.99$: the pressure profile in $uv$ space
}

\author{R. Gobat\inst{1}
\and E. Daddi\inst{2}
\and R.T. Coogan\inst{3}
\and A.M.C. Le Brun\inst{2}
\and F. Bournaud\inst{2}
\and J.-B. Melin\inst{2}
\and D.A. Riechers\inst{4,5,6}
\and M. Sargent\inst{3}
\and F. Valentino\inst{7}
\and H.S. Hwang\inst{8}
\and A. Finoguenov\inst{9}
\and V. Strazzullo\inst{10}
}

\institute{
Instituto de F\'{i}sica, Pontificia Universidad Cat\'{o}lica de Valpara\'{i}so, Casilla 4059, 
Valpara\'{i}so, Chile
\and AIM, CEA, CNRS, Universit\'{e} Paris-Saclay, Universit\'{e} Paris Diderot, Sorbonne 
Paris Cit\'{e}, F-91191 Gif-sur-Yvette, France
\and Astronomy Centre, Department of Physics and Astronomy, University of Sussex, Brighton, 
BN1 9QH, UK
\and Cornell University, Space Sciences Building, Ithaca, NY 14853, USA
\and Max-Planck-Institut f\"{u}r Astronomie, K\"{o}nigstuhl 17, D-69117 Heidelberg, Germany
\and Humboldt Fellow
\and Cosmic Dawn Center (DAWN), Niels Bohr Institute, University of Copenhagen, Juliane 
Maries Vej 30, DK-2100 Copenhagen; DTU-Space, Technical University of Denmark, Elektrovej 
327, DK-2800 Kgs. Lyngby, Denmark
\and Korea Astronomy and Space Science Institute, 776 Daedeokdae-ro, Yuseong-gu, Daejeon 
34055, Republic of Korea
\and Department of Physics, University of Helsinki, Gustaf H\"{a}llstr\"{o}pmin katu 2a, 
FI-00014 Helsinki, Finland
\and Department of Physics, Ludwig-Maximilians-Universit\"{a}t, Scheinerstr. 1, D-81679 
M\"{u}nchen, Germany
}

\date{}

\abstract{
We present Atacama Large Millimetre Array and Atacama Compact Array observations of the 
Sunyaev-Zel'dovich effect in the $z=2$ galaxy cluster Cl J1449+0856, an X-ray-detected 
progenitor of typical massive clusters in the present day Universe. 
While in a cleaned but otherwise untouched 92\,GHz map of this cluster, little to no negative 
signal is visible, careful subtraction of known sub-millimetre emitters in the \emph{uv} plane 
reveals a decrement at 5$\sigma$~significance. The total signal is $-190\pm36$\,$\mu$Jy, with 
a peak offset by 5''--9'' ($\sim$50\,kpc) from both the X-ray centroid and the still-forming 
brightest cluster galaxy. 
A comparison of the recovered \emph{uv}-amplitude profile of the decrement with different pressure 
models allows us to derive total mass constraints consistent with the 
$\sim6\times10^{13}$\,M$_{\odot}$~estimated from X-ray data. Moreover, we find no strong evidence 
for a deviation of the pressure profile with respect to local galaxy clusters, although a slight 
tension at small-to-intermediate spatial scales suggests a flattened central profile, opposite 
to what seen in a cool core and possibly an AGN-related effect. 
This analysis of the lowest mass single SZ detection so far illustrates the importance of 
interferometers when observing the SZ effect in high-redshift clusters, the cores of which 
cannot be considered quiescent, such that careful subtraction of galaxy emission is necessary.
}

\keywords{galaxies:clusters:individual:Cl J1449+0856 -- galaxies:clusters:intracluster medium}

\titlerunning{SZ observations of Cl1449}
\authorrunning{Gobat et al.}

\maketitle

\section{\label{hunger}Introduction}

The study of distant galaxy clusters has experienced a dramatic advance in the past decade, 
with the discovery of the first $z\sim2$~clusters \citep{And09,Gob11} and the 
subsequent breaching of that redshift limit into what was then considered then the epoch of 
protoclusters \citep{Spi12,Yua14,Wang16}.  It is now possible to efficiently identify galaxy 
clusters at $z>2$~in selected areas of the sky \citep[e.g.,][]{Chia14,Stra15,Dad17}, as 
well as select relatively large samples up to $z\lesssim2$~\citep[e.g.,][]{Wil13,Ble15}. 
We are thus leaving the discovery stage and are now becoming able to characterise the 
physical properties of these structures, with an eye toward answering longstanding questions 
regarding their baryonic content, such as the early evolution of their gaseous atmosphere 
(i.e., their intracluster medium, or ICM) and its interaction with their stellar component. 
The injection of energy into the ICM from star formation or active galactic nuclei (AGN) is 
a long-standing topic \citep{Kai91,Pon99,VS99,TN01}. However, high-redshift constraints are 
difficult to set, except indirectly in special cases \citep[e.g.,][]{Val16}, as both common 
methods for observing the ICM are less effective at higher redshift. X-ray observations, being 
limited by surface brightness, succumb to the inverse square law. The Sunyaev-Zel'dovich 
effect (SZ), on the other hand, is in principle distance-independent and has indeed yielded 
secure detections up to $z\sim2$~\citep{Bro12,Man14,Man18}. However, since the thermal SZ 
effect scales with electron density in the ICM, observations and surveys are still naturally 
biased towards massive ($\gtrsim10^{14}$\,M$_{\odot}$) systems. These not only become 
increasingly rare at higher redshift, but are also typically dominated by well-established 
quiescent galaxy populations \citep[e.g.,][]{Stan12,New14}, i.e., well past the stage where 
we would expect most of early energy injection to occur.\\

\noindent
Cl J1449+0856 (hereafter Cl1449) is a young galaxy cluster at $z=1.995$ \citep{Gob13} and 
one of the most distant with detectable X-ray emission. Serendipitously detected as an 
overdensity of red galaxies in \emph{Spitzer}/IRAC near-infrared imaging 
\citep[$m_{3.6}-m_{4.5}>0$;][]{Gob11}, it is a compact structure that already hosts a 
significant population of massive, quiescent galaxies \citep{Stra13,Stra16}, but also copious 
amounts of star formation as well as a $>100$\,kpc Ly$\alpha$~emission nebula in its core 
\citep{Val16}. The presence of a colder ($T\sim10^4$\,K) gas phase coexisting with the hot 
($T\sim10^6$\,K) ICM points to either a cool core \citep[e.g.,][]{Hec89}, which would be 
surprising at this early stage in the cluster's evolution, or feedback and maintenance from 
galactic outflows powered by either star formation or AGN~\citep{Val16}. 
In terms of mass, Cl1449 is a typical Coma progenitor at $z\sim2$~and therefore offers 
a window on the early thermodynamic evolution of typical galaxy clusters as well as the 
opportunity to study galaxy feedback to the ICM in a developing structure. 
We thus approach the SZ effect in this cluster from two different perspectives: as yielding 
an independent constraint on its total mass, providing a test for scaling relations at 
$z\sim2$~as well as a clearer picture of its place in galaxy cluster evolution, and as a 
probe of the thermodynamic status of its diffuse gas component.\\

\noindent
Here we present $\sim$92\,GHz observations of Cl1449 carried out with the Atacama Large 
Millimetre Array (ALMA) and the Atacama Compact Array (ACA), building upon recent work at 
millimetre and radio wavelengths \citep[hereafter S18 and C18, respectively]{Stra18,Coo18,Coo19}.  
We describe the observations in Section~\ref{ingredients}, the analysis of the data in 
Section~\ref{cooking}, and discuss its implication in Section~\ref{coffee}, while 
Section~\ref{burp} summarises our findings. We assume a $\Lambda$CDM cosmology with 
$H_0=70$\,km\,s$^{-1}$\,Mpc$^{-1}$, $\Omega_{\text{M}}=0.3$, and $\Omega_{\Lambda}=0.7$ 
throughout. Stellar masses and star formation rates (SFR) assume a \citet{Sal55} initial 
mass function.\\

\section{\label{ingredients}Observations and data reduction}

\begin{figure}
\centering
\includegraphics[width=0.49\textwidth]{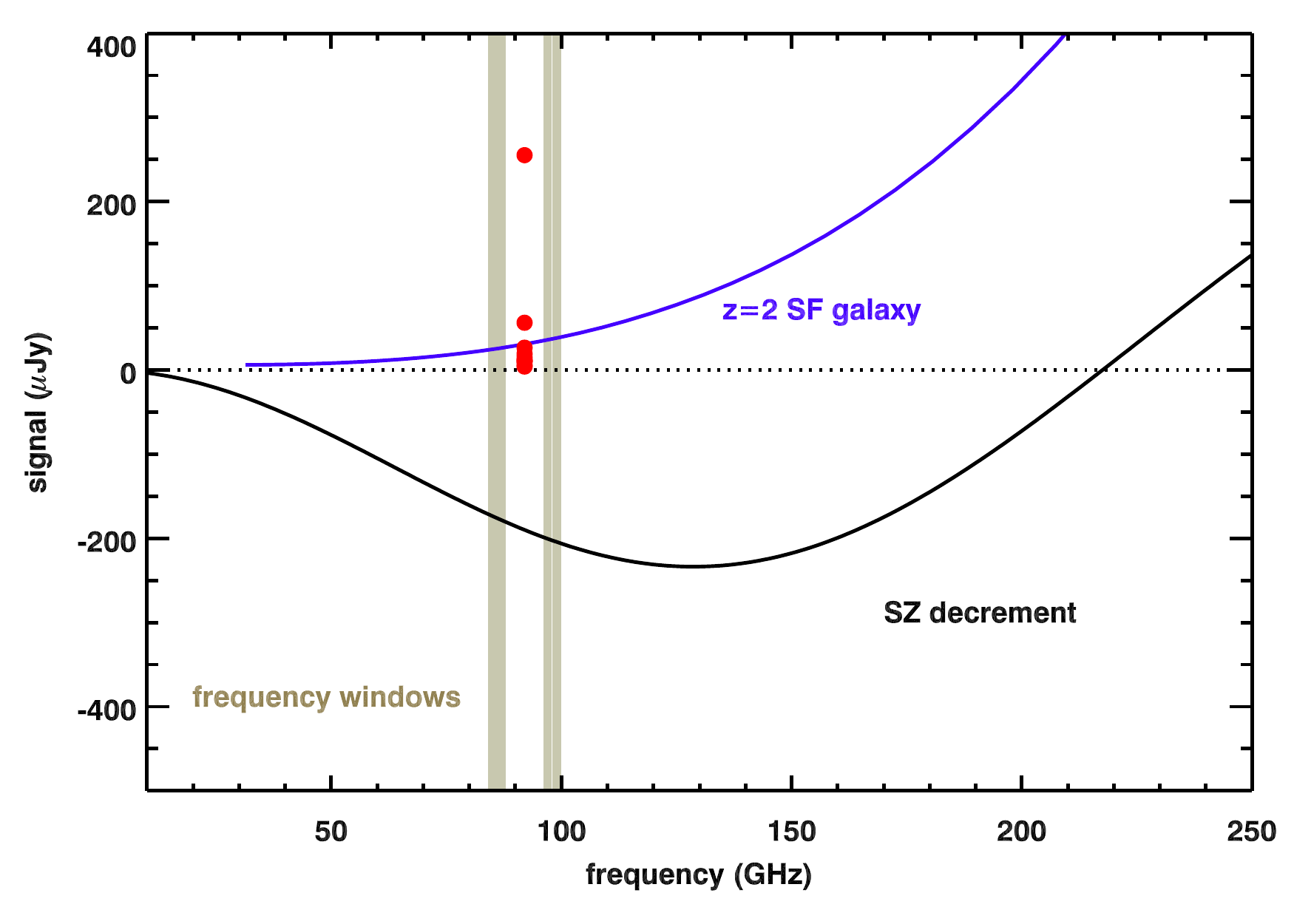}
\caption{SZ decrement as a function of frequency, scaled to our observed 92\,GHz value 
(see Section~\ref{cooking}), with a MS SED at $z=2$~for comparison 
\citep[blue curve;][]{Mag12}. The red dots show the 92\,GHz fluxes of point sources within 
the field of Cl1449, while the dark tan bands mark the frequency windows of our observations 
(lower and upper sub-band, respectively).
}
\label{fig:decrement}
\end{figure}

Cl1449 was observed with ALMA and ACA in Cycle 4 under program 2016.1.01107 (PI Gobat). 
The observations, which are summarised in Table~\ref{tab:obs}, were carried out between 
November 2016 and March 2017 as single pointings with total observing times of 49h for 
ACA and 9.7h for ALMA. The data were taken in Band 3, with a central frequency of 92\,GHz 
and a phase centre at R.A. = 14:49:14 and Dec = 8:56:26. 
Although not probing the peak of the SZ decrement, this frequency was chosen as a compromise 
to both optimise the total integration time and minimise positive contamination by the 
redshifted far-infrared emission from cold dust in star forming cluster members or 
high-redshift interlopers (Fig.~\ref{fig:decrement}). 
Our target of interest being extended, possibly over a scale of several tens of arcseconds, 
we chose the most compact ALMA configuration to minimise signal loss due to over-resolution 
(the maximum recoverable scale being 29'' in cycle 4) and probe large spatial scales. This is 
aided by our choice of frequency, generating the widest beam currently possible for both ALMA 
and ACA. As a result, the ACA and ALMA maps have 
synthesised beams of $\text{FWHM}_{\text{ACA}}\sim16.86''\times13''$~and 
$\text{FWHM}_{\text{ALMA}}\sim4.23''\times3.58''$, respectively, with a r.m.s. (root mean 
square) point-source sensitivity of $\sim$22 and $\sim$4\,$\mu$Jy/beam, respectively.\\ 

\begin{figure*}
\centering
\hspace{0.87cm}\includegraphics[width=0.417\textwidth]{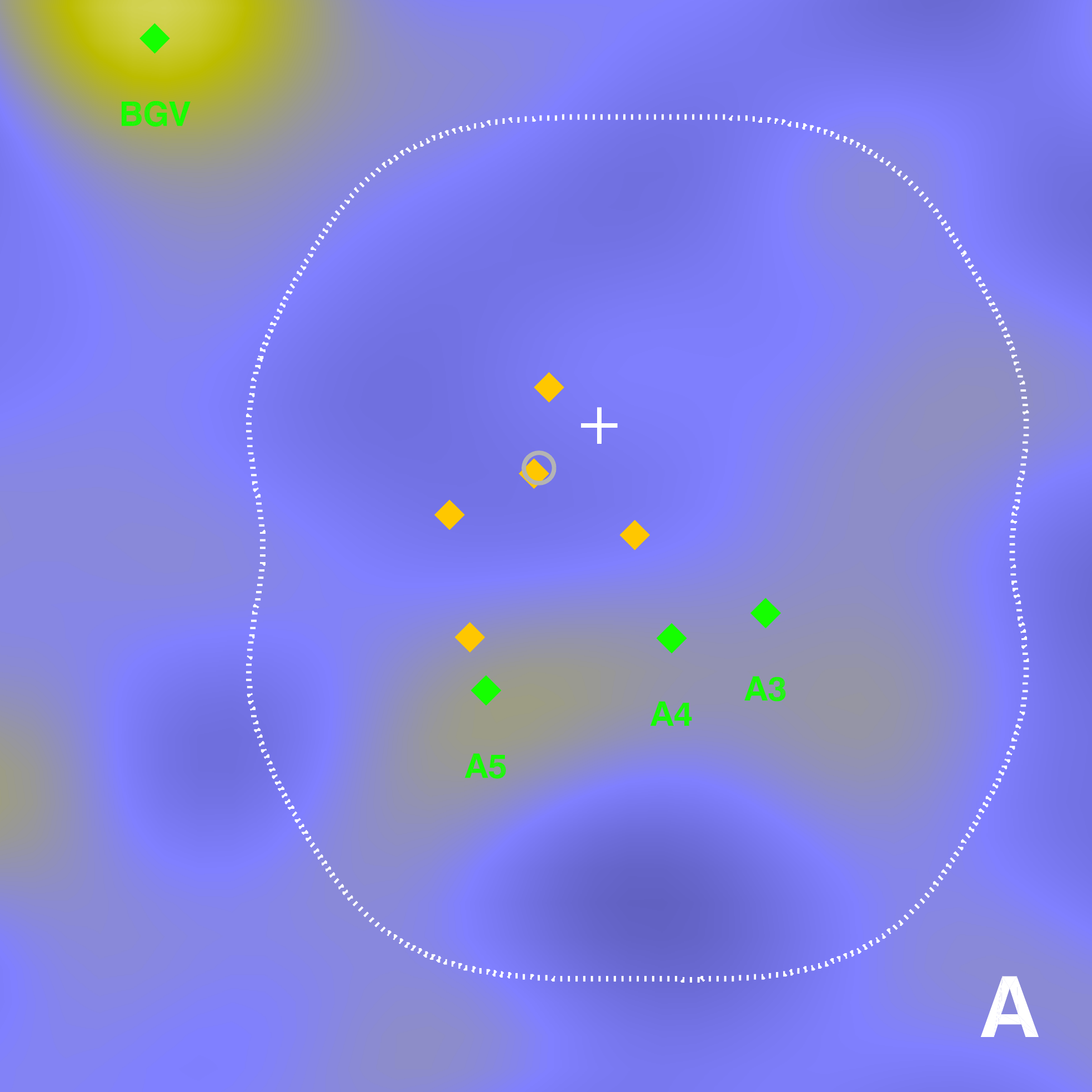}
\hspace{0.4cm}\includegraphics[width=0.417\textwidth]{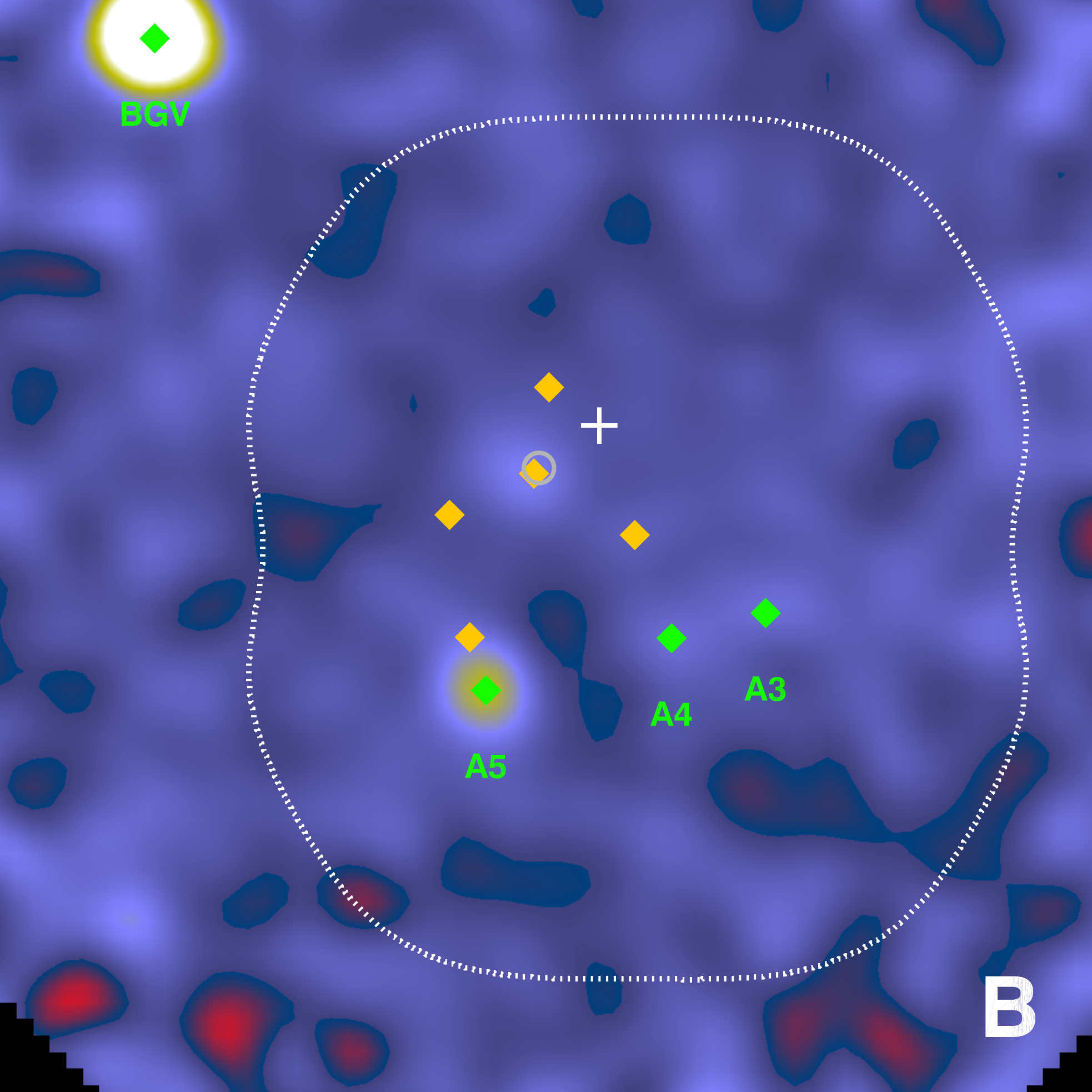}
\includegraphics[width=0.49\textwidth]{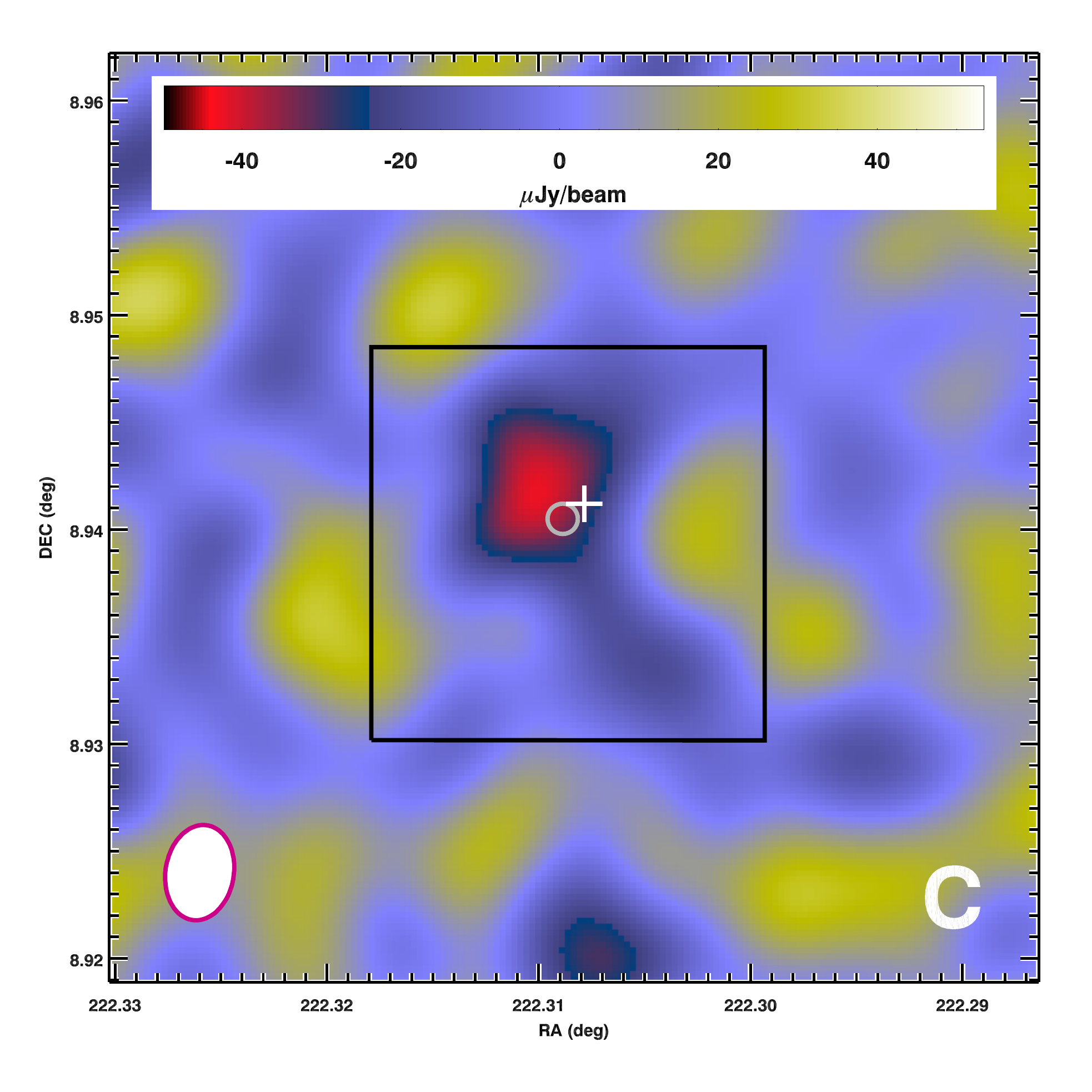}
\raisebox{0.87cm}{\includegraphics[width=0.417\textwidth]{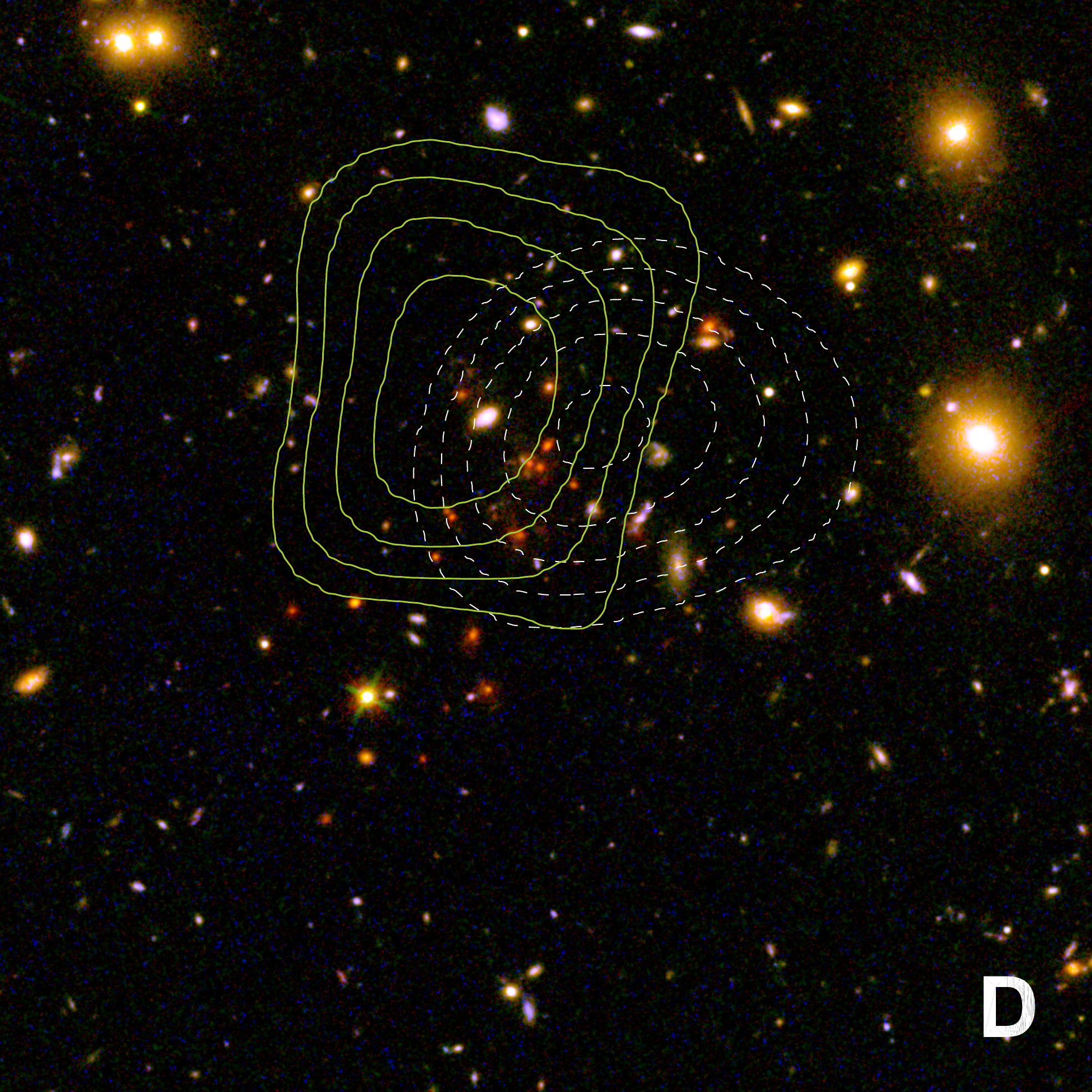}}
\caption{\emph{Top:} ACA (A) and ALMA (B) 92\,GHz maps of the field of Cl1449 
before point-source subtraction, created from the data using the \textsc{CASA} routine 
\texttt{CLEAN}. The white cross marks the centroid position of the extended X-ray emission 
seen by \emph{Chandra} while the grey circle shows the positions of the still-forming 
central galaxy of the cluster. The positions of all subtracted point sources are shown 
by diamonds (orange for confirmed cluster members, green with labels for confirmed or 
possible interlopers). For comparison, the dotted white contour marks the extent of the 
ALMA observations described in C18 and S18.
\emph{Bottom:} combined ACA+ALMA 92\,GHz image after subtracting point sources (C), 
showing the SZ signal from the cluster's ICM. The black square shows the field of 
view of panels A, B, and D, and the white-filled magenta ellipse the average synthesised 
beam size.
Panel D shows a \emph{HST}/WFC3 colour composite (F105W, F140W, and F160W) 
image of Cl1449 for comparison. The dashed grey contours display the X-ray emission 
as seen by \emph{Chandra}, while the light green contours shows the SZ signal above 
the r.m.s. noise.
}
\label{fig:image}
\end{figure*}

\noindent
We reduce the raw data using the \textsc{CASA} software suite \citep{CASA} and the script 
provided by ALMA to generate measurement sets (one per spectral window per array), which 
were merged into a single \textsc{UVFITS} table per array, for subsequent analysis with the 
\textsc{GILDAS}\footnote{\texttt{http://www.iram.fr/IRAMFR/GILDAS}} software suite. 
We use natural weighting for imaging throughout the paper. 
These data show, at first glance, little to no SZ signal (Fig.~\ref{fig:image}, A and B). 
This is not surprising as the field of Cl1449 is overdense in FIR sources, both 
within the cluster and in the background \citep[S18;][]{Smi19}. Despite our choosing a low 
frequency to mitigate the problem, the combined flux of high-redshift dusty sources is 
thus sufficient to fill the SZ decrement. We therefore subtract, from the data, point 
sources at the positions of 9 known FIR emitters (Table~\ref{tab:sources}). This is done 
on visibilities, i.e., in $(u,v)$~space. To determine the positions and fluxes of the sources, 
we use higher-resolution ALMA 870\,$\mu$m and CO(4--3) observations of Cl1449 (described in 
C18). We first measure their fluxes in the higher-resolution 92\,GHz ALMA data, using only 
visibilities with \emph{uv}-distance ($\sqrt{u^2+v^2}$; hereafter \emph{uv}) of $uv>30$\,m, 
i.e., considering only small spatial scales. These fluxes do not change significantly if we 
adopt a more stringent cut, such as $uv>50$\,m (corresponding to $\sim$100~kpc).
The sources are then subtracted from both the ALMA and ACA 92\,GHz data (this time over 
the whole \emph{uv} range), at the same fixed positions. In both cases we model them as 
point sources, since the beams are large compared to the sizes found in C18. Where 
possible we model and subtract the sources by groups of 4, iteratively from brightest 
to faintest, to minimise contamination. As a sanity check, we also compare the recovered 
fluxes to the 92\,GHz expectations from \citet{Mag12} spectral energy distribution (SED) 
templates, finding consistency (Fig.~\ref{fig:92ghz}).
A merged ALMA+ACA map of the resulting data is shown in Fig.~\ref{fig:image} (C), 
which shows a noticeable negative signal. Since only galaxies detected in either ALMA 
continuum or line emission maps were subtracted from the ALMA and ACA observations, 
some residual positive signal from below-threshold faint and/or low-mass galaxies might be 
still present in the data. The amplitude, and significance thereof, quantified in Section 
\ref{cooking} can thus be considered as conservative.\\

\begin{table}
\caption{Summary of ALMA and ACA observations of Cl1449}
\centering
\begin{tabular}{c c c c}
\hline\hline
& total time (h) & r.m.s. ($\mu$Jy/beam) & beam size ('')\\
\hline
ALMA & 9.7 & 4 & $4.23\times3.58$\\
ACA & 49 & 22 & $16.86\times13$\\
\hline
\end{tabular}
\label{tab:obs}
\end{table}

\section{\label{cooking}Analysis}

For both the ALMA and ACA data we extract fluxes by fitting, in bins of \emph{uv}, the 
complex visibilities with a point source using the \texttt{UV\_FIT} task \textsc{GILDAS}. 
We choose the point-source model for simplicity, as it corresponds in $(u,v)$~space to 
$V(u,v)=constant\times e^{-2\pi i(ux+vy)}$, where $(x,y)$~is the position of the source 
with respect to the phase centre and $(u,v)$~are here in units of cycles/distance. 
At phase centre, this is equivalent to averaging visibility amplitudes within the chosen 
\emph{uv} range.
However, we give $(u,v)$~in metres throughout the text for convenience, using the central 
frequency to convert these coordinates into distances.
We first perform the fit in a large bin of $uv=6-25$\,m, leaving the offset $(x,y)$ free 
to determine the location of the peak of the signal, then extract fluxes at fixed position 
in \emph{uv} bins of 6--13.5\,m, 13.5--18.75\,m, 18.75--25.5\,m, 25.5--50\,m, 50--100\,m, 
and 100--300\,m. We do not take the formal errors on the point-source fit as uncertainties 
to the signal in each \emph{uv} bin, but instead use the r.m.s. noise as measured in each 
bin with a point-source fit at randomised large offset positions. This yields slightly 
larger error bars on average.
The resulting \emph{uv}-amplitude profile, shown in Fig.~\ref{fig:modfit}, displays a 
negative signal in the $uv\sim6-30$\,m range, i.e., at angular scales $\gtrsim200$\,kpc. 
When combining ALMA and ACA visibilities, the total signal over all angular scales is 
$-190\pm36$\,$\mu$Jy, with a significance of 5.3$\sigma$. 
Including the errors on the fluxes of subtracted sources, weighted by their positions 
with respect to the best-fit pressure model (see Section~\ref{coffee}), would 
conservatively add another $\sim$4\,$\mu$Jy in quadrature to the uncertainty, which 
does not strongly affect the level of significance of the SZ detection.
Prior to the subtraction of positive sources, on the other hand, the SZ signal is only 
$\sim-20$\,$\mu$Jy, i.e., is almost entirely filled, with only the shortest \emph{uv} 
distances providing any tentative hint of a SZ decrement. 
If, on the other hand, we only remove sources which are either known to be interlopers 
or have not been conclusively proven to be at the cluster's redshift (i.e., A3, A4, A5, 
and BRG in Table~\ref{tab:sources}), a signal is marginally detected at 
$-123\pm40$\,$\mu$Jy. 
The filling of the decrement by confirmed cluster members thus amounts to $\sim$35\% 
of the signal, possibly more if either one of the unconfirmed sources (A4 and A5) is 
associated with the cluster. Assuming that Cl1449 is representative of its halo mass 
range and redshift, this test can be understood as an ideal unsubtracted case where no 
bright FIR interlopers are present along the cluster's line of sight. However, we 
currently have no reason to think that the field of Cl1449 is particularly overdense 
in FIR sources with respect to other, as-yet undiscovered clusters of its size and 
epoch. 
We also note a slight tension between the ALMA and ACA profiles within their range of 
overlap (Fig.~\ref{fig:modfit}, B), with the latter showing more decrement than the 
former. This cannot in principle be explained by differences between the two instruments, 
as simulated observations (see Section~\ref{chewing}) would rather suggest an opposite 
trend than the one observed, and might perhaps be due to a calibration issue. On the 
other hand, the significance of this difference is small enough ($\sim$1.5$\sigma$) that 
it can safely be attributed to noise.\\

\noindent
Cl1449 had previously been observed at 31\,GHz with the Combined Array for Research in 
Millimeter-wave Astronomy (CARMA). This observation, which has a r.m.s. noise of 
90\,$\mu$Jy/beam, did not yield a detection aside from some positive emission. 
Here we revisit the CARMA data and perform a similar point-source subtraction 
(Appendix~\ref{appendix:carma}) as discussed above. We find a loose constraint for the 
SZ signal of $\gtrsim$\,-360\,$\mu$Jy at 3$\sigma$, which is certainly consistent 
with the expectation of $\sim$\,-32\,$\mu$Jy from the 92\,GHz data when assuming a 
standard spectral shape.
On the other hand, \citet{Man14} report a secure detection of the similar-redshift cluster 
XLSSU J021744.1-034536 (hereafter XLSSC 122) with the same instrument and central frequency, 
but a $\sim$68\% larger integration time, matching its much larger mass.\\

\noindent
The peak of the weighted-average ALMA+ACA signal is offset from the phase centre by 
$\Delta(RA,DEC)=(4.4,4.3)$'', which translates into a separation of 4.7'' from the 
forming brightest cluster galaxy (BCG) and 9.5'' with respect to the peak of the X-ray 
emission, i.e., the putative centre(s) of mass of the cluster (Fig.~\ref{fig:image}, panels 
C and D).
Interestingly, \citet{Man14} also report an offset between the SZ signal and the BCG of 
XLSSC 122, which is of comparable amplitude when accounting for the different beam sizes of 
both datasets. That detection is consistent across different observations \citep{Man18} and 
thus rather unlikely to be a product of noise. Offsets of this amplitude between either the 
BCG or X-ray peak and the SZ centroid are not unexpected, especially in clusters that are 
in a unrelaxed state \citep[e.g.,][]{Zha14} as we would expect Cl1449 to be given its 
relative youth, and are commonly observed in high-redshift clusters 
\citep[e.g.,][]{Bro16,Stra19}.
We perform a Monte Carlo simulation to estimate the significance of this offset (see 
Appendix~\ref{appendix:cumul}), subtracting the combined astrometric uncertainty of ACA/ALMA 
and \emph{Chandra} in quadrature. We find that the difference between the SZ peak and the BCG 
positions is well within the normal variation of the simulation, while the offset between the 
SZ and X-ray peaks falls within the top 1.5\% of realisations, corresponding to a significance 
of at most 2.4$\sigma$. We therefore still cannot discount the possibility that this observed 
discrepancy between the peaks of the SZ and X-ray signals is simply due to random noise.\\

\subsection{\label{chewing}Modelling}

To investigate the characteristics of the ICM in Cl1449 and link the observed SZ 
signal to actual physical properties of the cluster, such as total mass, we fit 
the \emph{uv}-amplitude profile extracted from the ALMA and ACA data to a range of 
models with freely varying amplitudes. 
We first consider several models of the electron pressure profile of the ICM based 
on a generalised Navarro-Frenk-White functional form \citep[GNFW;][]{Nag07} with fixed 
parameter values:\\
$-$~the theoretical median profiles from \citet[hereafter LB15]{LeB15} based on 
cosmological hydrodynamical simulations, with different levels of feedback from AGN. 
In that paper, they are referred to as \textsc{REF}, \textsc{AGN 8.0}, and 
\textsc{AGN 8.5}. The last two, as the names suggest, include a prescription for AGN 
feedback with increasing intensity, while the \textsc{REF} model does not.\\
$-$~the empirical \citet{Arn10} profile (hereafter A10), derived from local 
$>10^{14}$\,M$_{\odot}$~galaxy clusters. This is also the profile used by \citet{Man18} 
to fit the SZ signal of the $z=1.99\pm0.06$~cluster XLSSC 122 and thus allows for direct, 
easy comparison with both this study and the low-redshift universe.
For completeness, we also include the empirical profile from \citet[hereafter S16]{Say16}, 
which is based on A10 but with a different outer slope. We adopt $\beta_{GNFW}=6.13$~as 
given in that paper, but note that S16 also find a mass and redshift dependence to the 
slope $\beta_{GNFW}$~which, for Cl1449, would correspond to its A10 value.\\
$-$~the empirical ``high-z'' profile from \citet[hereafter McD14]{McD14}, which is based on 
a sample of $z=0.6-1.2$, $>$$10^{14}$~M$_{\odot}$ galaxy clusters observed with the South Pole 
Telescope. It differs from the A10 profile mainly by being flatter (i.e., having less pressure) 
at small radii. We consider both the cool core and non-cool core versions of this profile.\\

\begin{figure*}
\centering
\includegraphics[width=0.49\textwidth]{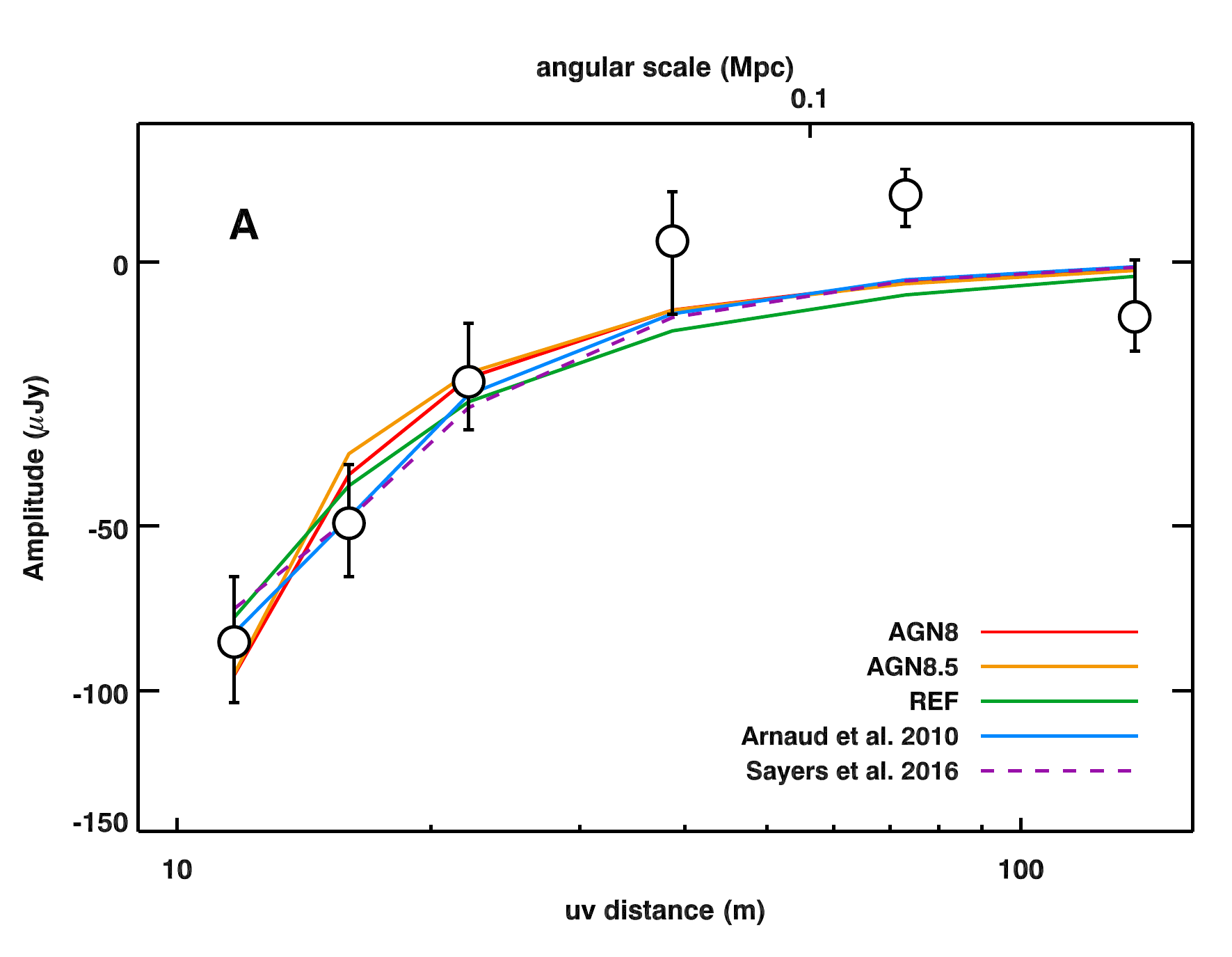}
\includegraphics[width=0.49\textwidth]{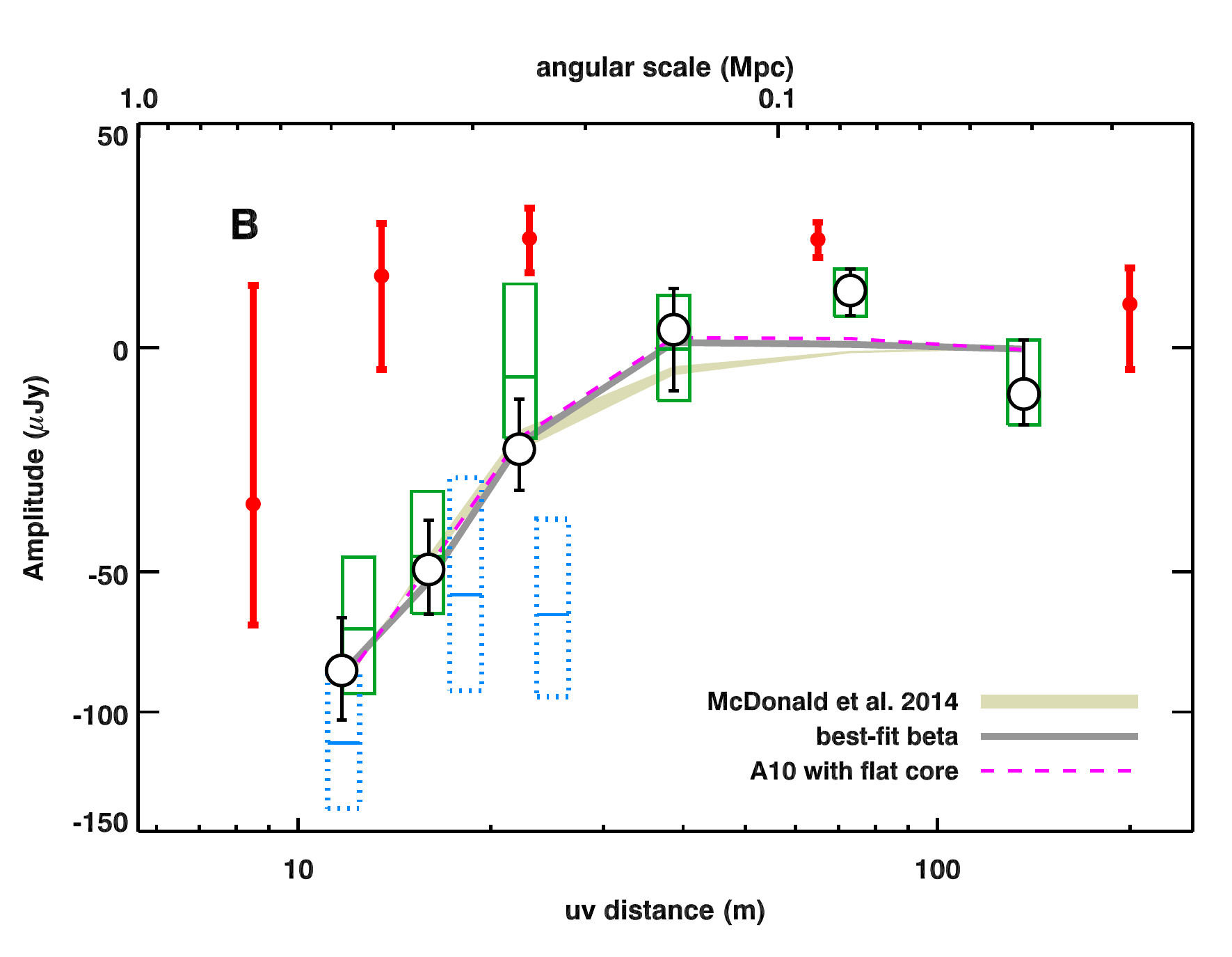}
\caption{\emph{A:} Amplitude of the SZ decrement as a function of 
baseline length and physical scale for the combined ALMA+ACA dataset (empty 
white circles with error bars), after subtraction of positive sources 
(active galaxies). 
For comparison, coloured lines show rescaled theoretical models with different 
feedback schemes from LB15 as well a rescaled empirical pressure profile of local 
clusters from A10, including the modified outer slope from S16.
\emph{B:} 
same data, with best-fit $\beta$-model. The solid green and dotted blue rectangles 
show the individual ALMA and ACA \emph{uv}-amplitude profile and noise, respectively, 
while the red points with error bars show the combined amplitudes \emph{before} 
subtraction of positive point sources. 
The tan curve shows the range covered by the McD14 best-fit models, while the magenta 
dashed one shows a composite model using the A10 profile at $>7$'' and a flat signal 
in the inner 7''.
}
\label{fig:modfit}
\end{figure*}

\noindent
For each model we create a map of the intrinsic signal by projecting the profile on 
the plane of the sky at the coordinates of the cluster, including an average 
Compton-\emph{y} background of $1.6\times10^{-6}$~derived from the 25\,deg$^2$~simulated 
maps described in LB15. For simplicity all models are spherical, i.e., axisymmetric when 
projected. 
We also consider the contribution of a gravitationally lensed infrared background to the 
decrement. However, we estimate it to be minute ($<0.1$\%; Appendix~\ref{appendix:lensedib}) 
due to a combination of low halo mass, decreasing lensing efficiency at higher redshift, 
and low 92~GHz background density.
We use the 2D models as inputs for noise-free simulations of ACA and ALMA 
observations using the \texttt{simalma} task in \textsc{CASA}, taking care to adopt the 
same integration times and hour angles as with the data. We then compare the model and 
data visibilities, merging the ALMA and ACA deviates as the last step. 
While we keep the GNFW parameters fixed to the various models' values, we let the mass 
vary freely. However, while the models are given in function of $M_{500}$, our previous 
works \citep{Gob11,Val16} instead discuss the ``total'' mass $M_{200}$. We here therefore 
extrapolate $M_{200}$~from $M_{500}$, assuming that the mass distribution follows the 
GNFW profile. The fits of the A10, LB15, S16, and McD14 models to the observed 
\emph{uv}-amplitude profile are shown in Fig.~\ref{fig:modfit}.\\

\noindent
Additionally, we also attempt a parametric fit using the simpler $\beta$-model 
historically used to describe the X-ray luminosity profiles of galaxy clusters 
\citep{CF78}, leaving both the core radius $r_c$~and the index $\beta$~free to vary. 
For practicality, in this case we fit the ALMA and ACA visibilities with the forward 
Fourier transform of each $\beta$-model sampled at the same $(u,v)$~positions as the 
data. At our signal-to-noise ratio (S/N) this is essentially equivalent to the full 
\texttt{simalma} model (see Appendix~\ref{appendix:fft}), and allows us to explore 
the parameter space of the models more rapidly and at little to no detriment to 
precision.\\

\section{\label{coffee}Results and discussion}

\begin{figure}
\centering
\includegraphics[width=0.5\textwidth]{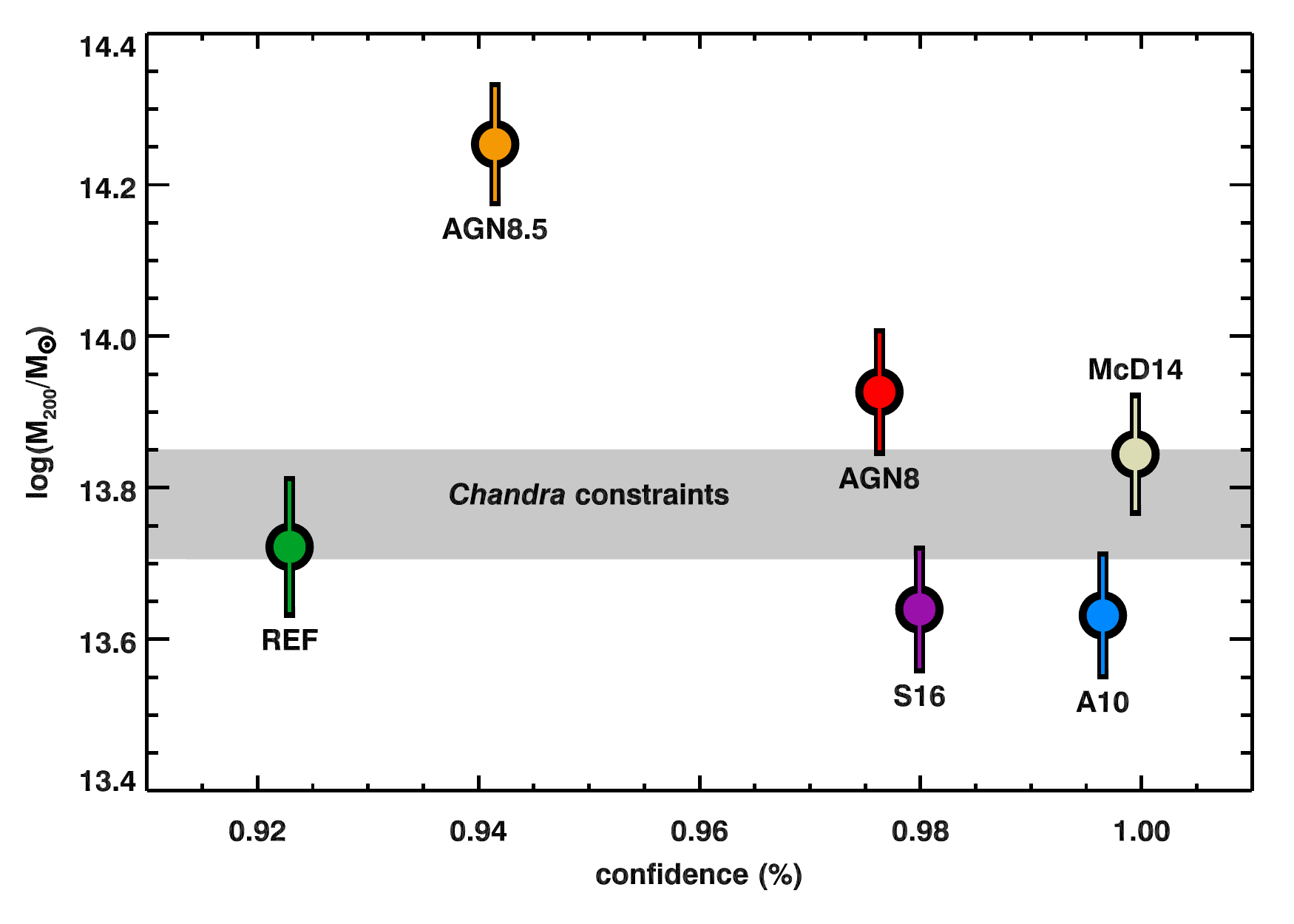}
\caption{Total mass constraints obtained from fitting the observed 
uv-amplitude profile with theoretical/empirical models (Fig.~\ref{fig:modfit}, A), 
as a function of the confidence of the fit derived from the $\chi^2$. The grey 
horizontal stripe shows the 1$\sigma$~\emph{Chandra} mass constraint, for comparison.
}
\label{fig:modmass}
\end{figure}

\begin{figure}[!h]
\centering
\includegraphics[width=0.49\textwidth]{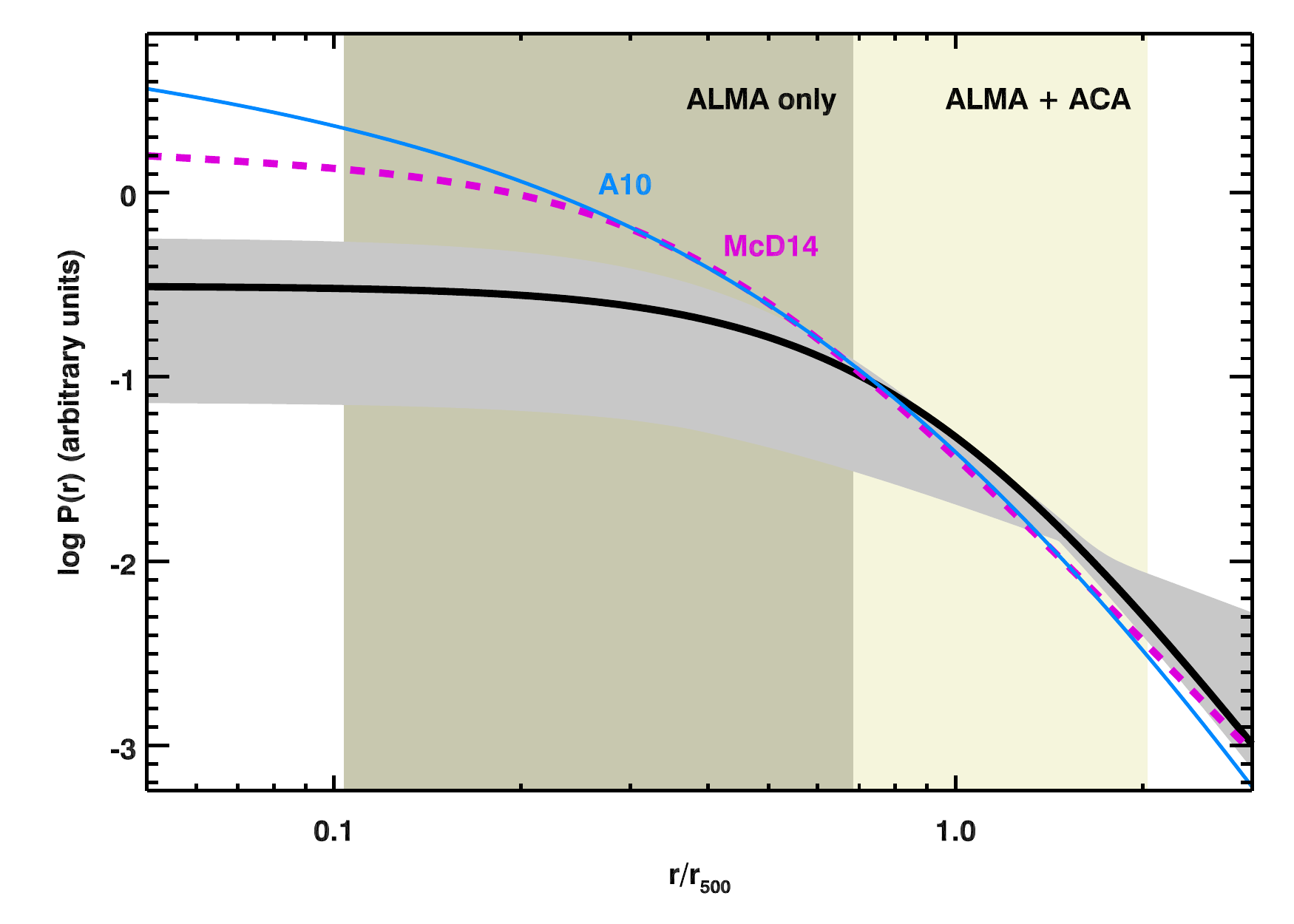}
\caption{Best-fit $\beta$~model (black curve) and 1$\sigma$~confidence envelope 
(grey). For comparison, the A10 and average McD14 models are shown in blue and magenta, 
respectively, and the spatial range probed by ALMA and ACA by light and dark tan regions.}
\label{fig:betaprofile}
\end{figure}

We find that the McD14 profile matches the observed data best, as determined by its 
$\chi^2$~value, followed by A10 and the AGN-feedback LB15 models with only the 
\textsc{REF} one falling below the $\sim$2$\sigma$~confidence level. Most suggest total 
cluster masses that are consistent with the \emph{Chandra} constraint of 
$M_{200}\sim6\times10^{13}$\,M$_{\odot}$~\citep{Val16} (Fig.~\ref{fig:modmass}), i.e., 
at least a factor $\sim$2 below the limit of typical SZ surveys at any redshift 
\citep{Ble15,P27,Hil18}. Among the various models considered, only \textsc{AGN 8.5} 
yields a higher mass of $\sim$$1.8\times10^{14}$\,M$_{\odot}$. 
In this case, the somewhat higher value is unsurprising since, in the models, 
the gas fraction decreases with increasing AGN feedback as more material is ejected, 
thereby requiring a higher mass to reproduce the same integrated signal. 
Overall, the constraining power of the observed \emph{uv}-amplitude profile with 
respect to the pressure model is somewhat limited, especially at large scales/small 
baselines where the GNFW models appear to be equivalent to one another, in part due to 
the relatively modest S/N of the data. Of the fixed-parameter models, only the McD14 
one fits noticeably better, at intermediate \emph{uv}.
We also note that the profile can be reproduced best with a $\beta$-model 
(Fig.~\ref{fig:modfit}, B), which is unsurprising as it has two additional free 
parameters. However, owing to parameter degeneracy (see Appendix~\ref{appendix:beta}, 
Fig.~\ref{fig:betacontours}), the constraints it provides remain loose as well, with 
$(r_c,\beta) \gtrsim (100 \text{ kpc},0.4)$. 
Nevertheless, we note that the core radius $r_c$~is consistently large, of the order 
of the (putative, X-ray derived) $r_{500}$~of the cluster whereas, by comparison, the 
galaxy density profile has $r_c\sim20$\,kpc \citep{Stra13}. 
Consequently, the best-fitting profiles are essentially flat at $\lesssim0.3 r_{500}$ 
(i.e., in the inner $\sim150$\,kpc; Fig.~\ref{fig:betaprofile}).\\

\noindent
This is due to the apparent lack of power of the observed profile at $uv\sim30$\,m 
(corresponding to $\lesssim10$''), which only the McD14 profile reproduces at all 
\emph{uv} within the uncertainties of the data (the non-cool core and cool core versions 
of the profile show here little difference; Fig.~\ref{fig:modfit}, B). The other GNFW 
models only fit completely if we force the projected signal to be constant (i.e., flat) 
within the inner $\sim$15'' (see Fig.~\ref{fig:modfit}, B).
Assuming for the sake of speculation that this flatness at intermediate and large 
\emph{uv} distances is not simply due to noise fluctuations, as suggested by the evolution 
observed in higher-mass (but lower-redshift) clusters \citep[McD14]{Bat12}, at least two 
different causes can be envisioned. 
On the one hand, left-over flux from incomplete point source subtractions could indeed 
remove power from the SZ signal at small scales. This might either arise from an 
underestimate of the (here, 92\,GHz) flux of detected star-forming sources or from the 
emission of galaxies below the detection threshold. In the second case, this would 
correspond for this cluster to an additional flux of $\sim$10\,$\mu$Jy, or about 
$\sim$60\,M$_{\odot}$\,yr$^{-1}$~for main sequence (MS) galaxies at the cluster redshift 
\citep{Mag12}. This would represent $\sim$8\% of the SFR within the region covered by the 
SZ signal (C18), considering cluster members and interlopers. 
For comparison, this corresponds to three $\sim$$10^{10}$\,M$_{\odot}$~MS galaxies at 
the redshift of the cluster or 30--40 $10^9$\,M$_{\odot}$~galaxies \citep{Sch15}. 
The mass completeness limit of our deepest near-infrared imaging being
$\sim$10$^{10}$\,M$_{\odot}$~\citep{Stra13}, it is not impossible that a few galaxies 
might have been missed even in the priors catalogue. 
Diffuse emission could also provide another source of positive signal at slightly larger 
scales. In addition to its hot ICM and giant Ly$\alpha$ nebula, Cl1449 also hosts 
intracluster light (ICL), on a similar scale to the Ly$\alpha$ emission and possibly of 
stellar origin (Dimauro et al., in prep.). Thermal emission from intracluster dust might 
have been detected at lower redshift 
\citep[however, the low resolution of the data makes it unclear;][]{P43}, although constraints 
on the gas-to-dust ratio of the ICM place it at a much lower level than in star forming 
galaxies \citep{Kit09,Gut17}. However, since Cl1449 is in a much younger dynamical state, 
as also evidenced by its relatively bright ICL, its FIR emission could be comparatively 
higher.\\

\noindent
On the other hand, if the lack of negative power at $uv>30$\,m is an intrinsic property of 
the SZ signal, it suggests lower central electron density and/or temperature with respect 
to lower-redshift clusters. This could either simply reflect a secular evolution in clusters' 
ICM pressure distribution or be the result of feedback effects from galaxies. In the latter 
case, AGN-generated cavities in the ICM, for example, typically have lower electron density 
and pressure than the thermal ICM, leading to a decreased signal with respect to the thermal 
case \citep[e.g.,][]{Pfr05,Ehl19,Abd19}. 
It would not be entirely surprising for one to be present in Cl1449, as the cluster hosts 
at least two X-ray detected AGNs, whose putative outflows are likely associated with the 
powering and/or maintenance of the Ly$\alpha$~emission nebula in its core \citep{Val16}.
Furthermore, while the extent of the ``flat'' pressure region necessary to reproduce the 
observed profile is large, of the order of $\sim$100\,kpc, it is not unheard of in clusters 
\citep[e.g.,][]{Abd19}. One might therefore find it puzzling that the McD14 and A10 models, 
which assume no baryonic physics, match the observed profile and X-ray mass constraint better 
than the AGN 8.0 and AGN 8.5 models, which include them. 
Additionally, these models were calibrated on $z\sim0$~data and assume self-similar 
evolution with redshift, as do the scaling relations \citep{Lea10} used in the 
\emph{Chandra} analysis. Contrarily, more recent work suggest that the assumption of 
self-similarity does not quite hold when AGN feedback is considered \citep{LeB17}.\\

\begin{figure}[!h]
\centering
\includegraphics[width=0.49\textwidth]{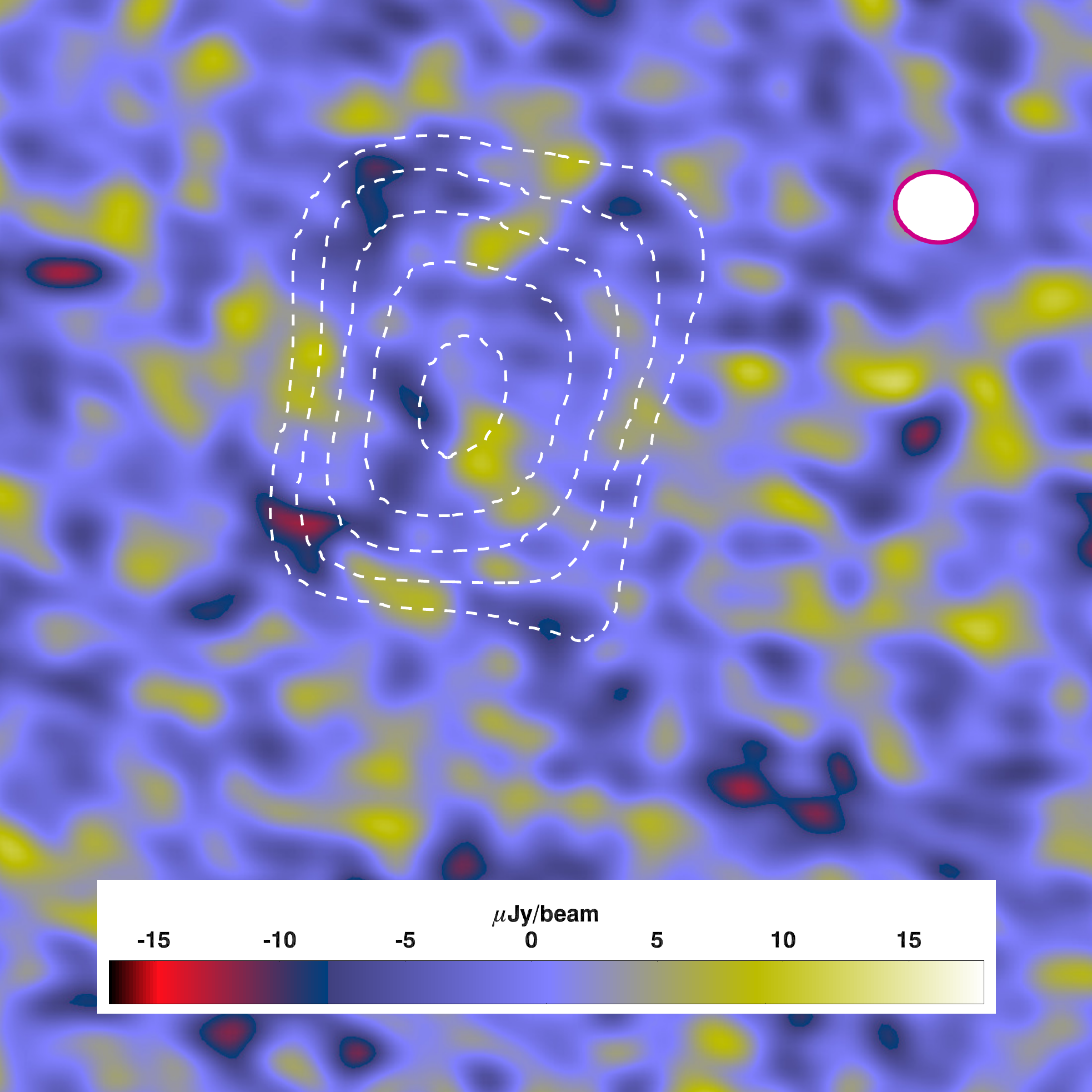}
\caption{ALMA 92\,GHz residual image, after subtracting both the point sources and the 
SZ signal, with the same field of view as panels A, B, and D of Fig.~\ref{fig:image}. 
For comparison, the white dashed contours mark the position of the SZ signal as shown 
in panels C and D of Fig.~\ref{fig:image}. The white-filled magenta ellipse shows the 
size of the synthesised ALMA beam.}
\label{fig:almaresid}
\end{figure}

\noindent
Finally, we note that a deviation from axisymmetry in the SZ signal, such as non-zero 
ellipticity, imply that either of the effects discussed above (or combination thereof) 
would be stronger, as it would transfer power to smaller scales, i.e., flatten the 
\emph{uv}-amplitude profile. The excellent agreement between the observed profile 
and models shown in Fig.~\ref{fig:modfit} (B) suggests however that the SZ decrement 
of the cluster has a fairly circular geometry. On the other hand, no elliptical or 
multi-component fit was attempted given the S/N of the data. Even with ALMA, we are, 
in the $z\sim2$~regime, probing the limits of the recoverable information. 
The r.m.s. noise of the ALMA data and the lack of detectable structure in its residuals 
after subtracting both the point sources and the SZ signal (as shown in 
Fig.~\ref{fig:almaresid}) allows us to put a 3$\sigma$ upper-limit on individual 
inhomogeneities in the SZ signal of $\sim$6\% of the total decrement. However, 
lower-amplitude pressure discontinuities might still be present. 
The current data nevertheless provide an interesting baseline for comparison with 
future observations of similar or higher-redshift galaxy clusters, such as Cl J1001+0220 
at $z=2.5$~\citep{Wang16}, in which feedback from highly-active galaxies is expected 
to be strong. Conversely, averaging the SZ signal over a population of high-redshift 
galaxy clusters, by increasing the S/N and minimising cosmic variance, would allow us 
to set true constraints on ICM pressure models at early stages of cluster formation.\\

\section{\label{burp}Conclusions}

Combined ALMA and ACA observations of Cl1449 at 92\,GHz have yielded a secure 
$\sim$5$\sigma$~detection of the SZ decrement associated with its ICM.
Comparing the \emph{uv}-amplitude profile of the SZ signal to a variety of pressure 
models, we confirm the total mass estimates obtained from \emph{Chandra} X-ray 
observations of the cluster. While the SZ signal provides independent constraints, 
these still depend on the adopted model and its calibration. We find a factor $\sim$2 
spread in mass estimates among models at similar significance levels, with the SZ 
constraints nevertheless clustering around the mass inferred from \emph{Chandra} X-ray 
data.\\

\noindent
In this work we measure the 92\,GHz flux of galaxies in the cluster's field and subtract 
it from the complex visibilities, i.e., in the Fourier space of the data. We perform 
the rest of the analysis entirely on the visibilities, rebinning them into a 
\emph{uv}-amplitude profile. The final S/N of the data, conservatively estimated, is 
not quite high enough to strongly constrain ICM pressure models. 
We see no sign of a cool core and, while the empirical $z<0.2$ \citet{Arn10} profile 
appears to hold on here as well, we notice a small tension between the data and 
locally-calibrated models. This could either be produced by residuals from the subtraction 
of positive sources or might reflect a pressure deficit in parts of the cluster's ICM 
compared to expectations, as suggested by the redshift trend seen in less distant and more 
massive clusters. Distinguishing between these two scenarios is not possible with the 
current data.\\

\noindent
The density of star formation present within Cl1449 is sufficient to almost entirely 
fill the SZ decrement unless corrected for. This issue is likely to affect all 
$z\gtrsim2$~clusters and to grow in severity with redshift due to both the 
increased activity of galaxies within cluster cores 
\citep[e.g.,][]{Wang16,Wang18} and the negative $K$-correction of their FIR dust 
emission at the frequencies of the SZ effect. It can nevertheless be slightly 
minimised by observing at lower frequencies, since at $z>2$ the tail of dust emission 
in galaxies falls somewhat steeper than the SZ decrement. 
For example, we would expect in the case of Cl1449 a $\sim$30\% improvement in contrast 
when observing with ALMA/ACA in Band 1 ($\sim$40\,GHz; not yet commissioned at the 
time of writing) instead of Band 3. At $z\sim2.5$, on the other hand, the gain would 
be closer to $\sim$80\%. Although the SZ signal in Band 1 is also expected to be lower 
by a third compared to that in Band 3, it will be sampled by a beam a factor $\sim$3 
larger. A simple calculation using our best fitting profile and the noise predictions 
from the current exposure time calculator then suggests that we can reach a comparable 
S/N at 40\,GHz with the same integration time as for 92\,GHz, but with considerably less 
uncertainty on the contamination from positive emitters.\\

\noindent
Our observation of the lowest mass single SZ detection so far demonstrates the power of 
ALMA for the study of the ICM of emerging galaxy clusters. 
It also illustrates the usefulness of combining short- and long-baseline interferometric 
observations in the context of SZ surveys. Indeed, the necessity for point-source 
subtraction, which requires a good prior knowledge of FIR emitters in the target field, 
as well as the increasing activity of cluster galaxies as we peer further back in time, 
casts doubts on the viability of single-dish telescopes for high-redshift SZ surveys. 
In this case multi-band observations would be absolutely necessary, as well as a high 
S/N to compensate for the steeply rising FIR SED of star forming galaxies, in both 
frequency and redshift, with respect to the SZ signal. 
However, this might still not be sufficient without accurate redshift information, 
as such observations could be susceptible to degeneracies between the spectral shape 
of the SZ signal and the line-of-sight distribution of FIR emitters. We can thus 
expect a significant and increasing number of structures to be misidentified or missed 
entirely due to star formation filling their decrement, suggesting that any SZ census 
of $z\sim2$~clusters is at risk of being biased towards older galaxy populations 
rather than simply higher relative total masses.\\

\begin{acknowledgements}

We thank the anonymous referee for their constructive for their constructive 
report, which helped improve the presentation of this work.
This work is based on data from ALMA program 2016.1.01107, as well as programs 
2012.1.00885.S and 2015.1.01355.S. ALMA is a partnership of ESO (representing its 
member states), NSF (USA) and NINS (Japan), together with NRC (Canada), MOST and ASIAA 
(Taiwan), and KASI (Republic of Korea), in cooperation with the Republic of Chile. 
Support for CARMA construction was derived from the G. and B. Moore Foundation, the 
K. T. and E. L. Norris Foundation, the Associates of the California Institute of 
Technology, the states of California, Illinois, and Maryland, and the NSF. Ongoing 
CARMA development and operations are supported by the NSF under a cooperative 
agreement, and by the CARMA partner universities. 
R.T.C. acknowledges support from the Science and Technology Facilities Council 
grant ST/N504452/1; F.V. acknowledges support from the Villum Fonden research grant 
13160 ''Gas to Stars, Stars to Dust: Tracing Star Formation across Cosmic Time; 
A.M.C.L.B. was supported by the European Research Council under the European Union's 
Seventh Framework Programme (FP7/2007-2013) / ERC grant agreement number 340519; 
D.R. acknowledges support from the National Science Foundation under grant number
AST-1614213 and from the Alexander von Humboldt Foundation and the Federal Ministry 
for Education and Research through a Humboldt Research Fellowship for Experienced 
Researchers.

\end{acknowledgements}

\begin{appendix}

\section{\label{appendix:subtracted}92\,GHz continuum sources}

\begin{figure}[!h]
\centering
\includegraphics[width=0.49\textwidth]{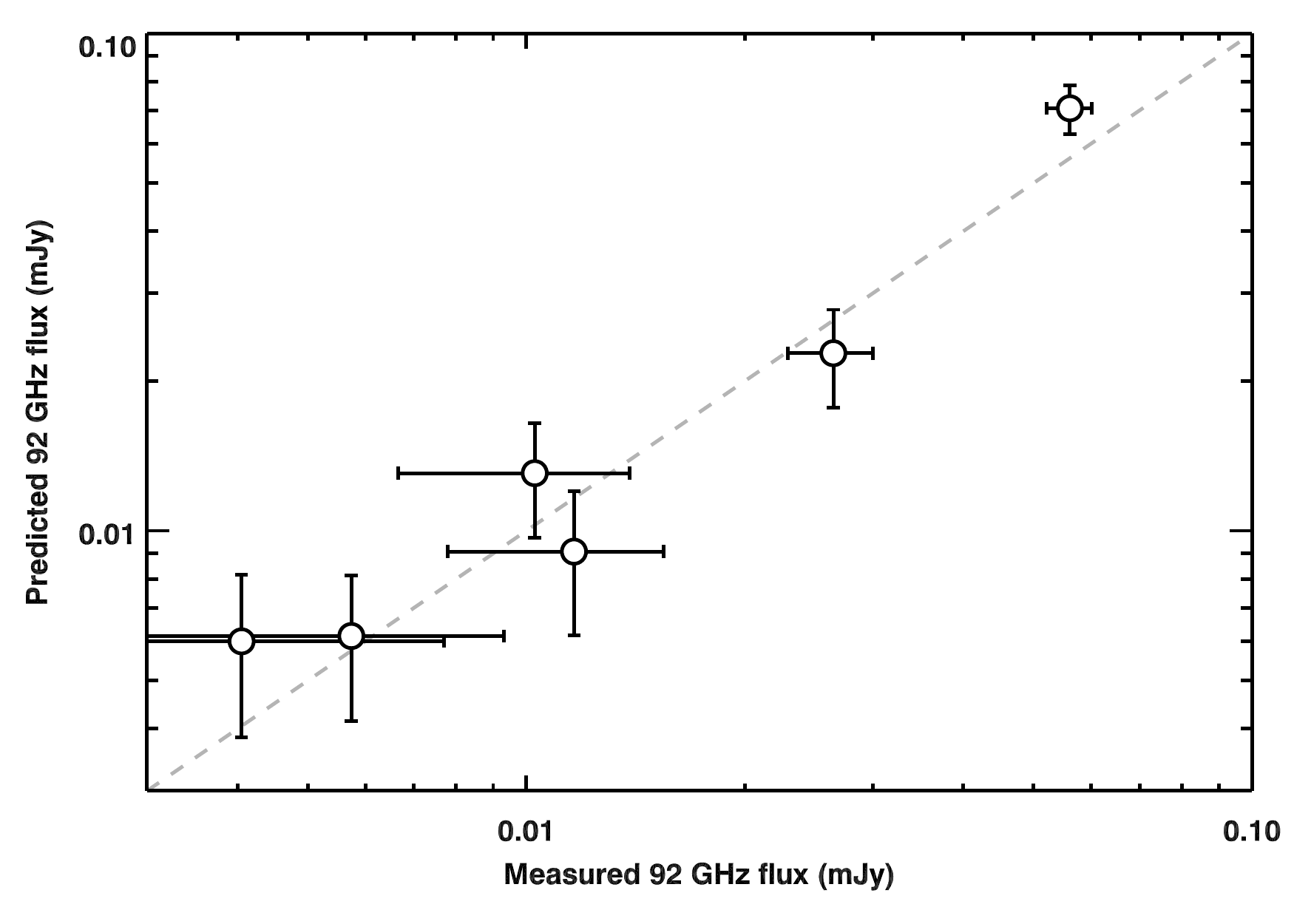}
\caption{
Comparison between the subtracted 92\,GHz fluxes of sources, i.e., measured from the 
ALMA data, and the fluxes predicted by \citet{Mag12} MS SED templates, based on the 
SFR of the sources derived in C18 from CO fluxes. Only sources for which a redshift 
estimate (photometric or spectroscopic) is available are shown.
}
\label{fig:92ghz}
\end{figure}

Table~\ref{tab:sources} shows the 92\,GHz fluxes of known FIR sources in the field 
of Cl1449, measured on our ALMA data. We compare these to the predictions of 
\citet{Mag12} templates (Fig.~\ref{fig:92ghz}), assuming MS SEDs based on the 
overall consistency between the MS at $z\sim2$~and the SFRs derived by C18.

\begin{table}[h]
\caption{Known 92\,GHz emitters in the field of Cl1449, after correcting for the 
primary beam}
\centering
\setlength{\tabcolsep}{8pt}
\begin{tabular}{c c c c c}
\hline\hline
ID$^*$ & R.A. & Dec & Flux  & Flux error\\
& deg & deg & $\mu$Jy & $\mu$Jy\\
\hline
A1+B1$^{**}$ & 222.30882 & 8.94054 & 26.5 & 3.6\\
A5 & 222.30963 & 8.93690 & 56.1 & 4.0\\
6$^{**}$ & 222.30991 & 8.93779 & 11.6 & 3.8\\
BRG & 222.31526 & 8.94785 & 255.2 & 8.5\\
A2$^{**}$ & 222.30710 & 8.93951 & 10.3 & 3.6\\
A3 & 222.30488 & 8.93820 & 14.6 & 4.0\\
A4 & 222.30648 & 8.93778 & 19.0 & 3.9\\
13$^{**}$ & 222.30856 & 8.94199 & 5.7 & 3.6\\
N7+S7$^{**}$ & 222.31025 & 8.93985 & 4.1 & 3.6\\
\hline
\end{tabular}
\justify
$^{*}$~identifiers in C18, except for \textsc{BRG} which denotes a bright, 
low-redshift radio galaxy outside the field of the data discussed in that paper. 
Given the lower resolution of our ALMA data, some close sources in C18 were merged 
for the purpose of 92\,GHz subtraction.
$^{**}$~confirmed cluster members.
\label{tab:sources}
\end{table}

\section{\label{appendix:cumul}Peak offset significance}

\begin{figure}[!h]
\centering
\includegraphics[width=0.49\textwidth]{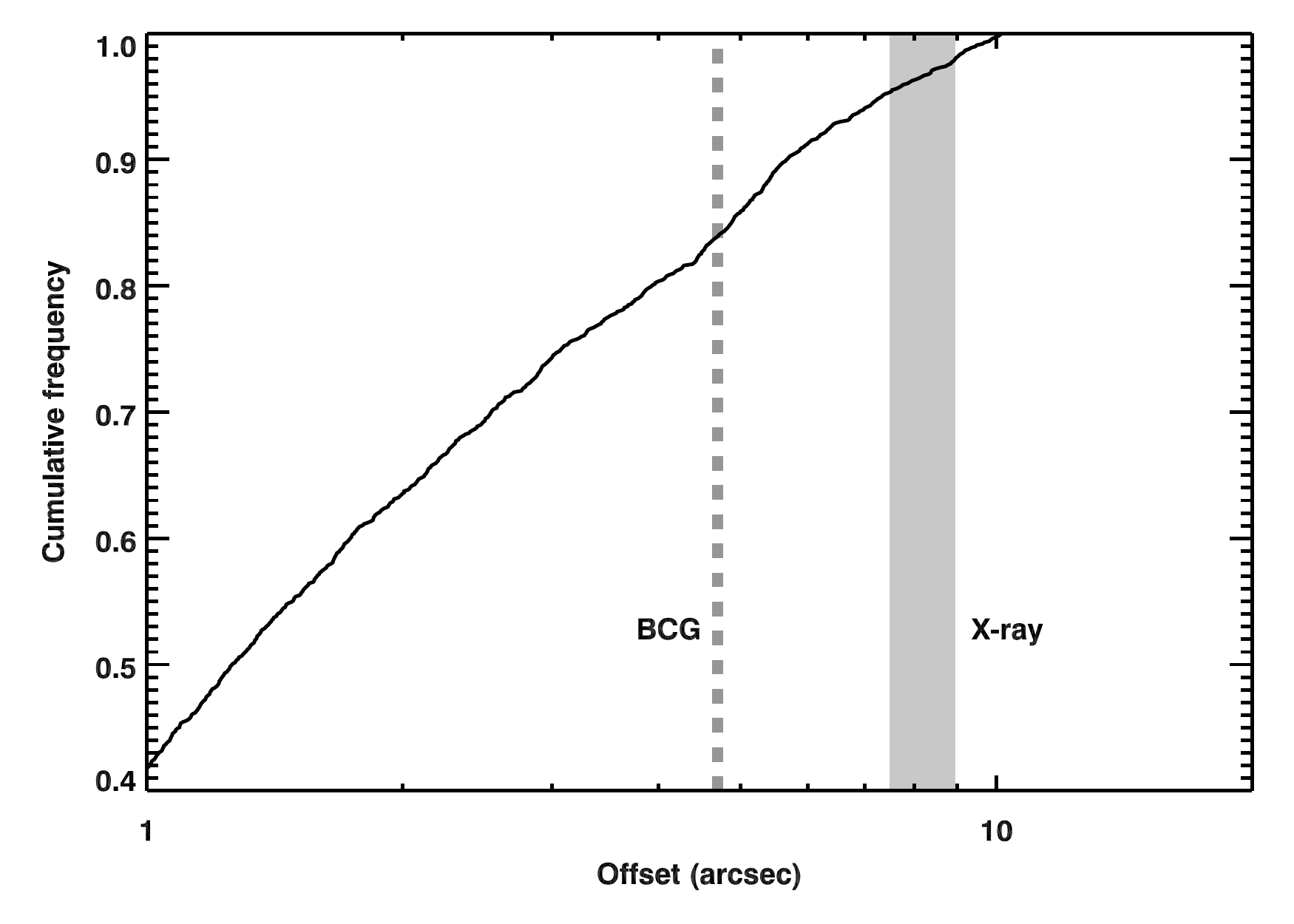}
\caption{
Cumulative distributions of the positional offset between the true and recovered 
peak of the SZ signal. These are based on Monte Carlo simulations using the $(u,v)$~sampling 
of the data visibilities, assuming a noise consistent with the observed one and using a 
\citet{Arn10} pressure model with a total mass of $6\times10^{13}$\,M$_{\odot}$. 
The grey band (dashed vertical line) shows the offset between the positions of the SZ and 
X-ray peaks (respectively, BCG).
}
\label{fig:offset}
\end{figure}

To investigate the significance of the observed offset between the positions of the 
peaks of the SZ and X-ray signals, we carry out a simple Monte Carlo simulation using a 
simulated SZ decrement based on the A10 pressure model and best-fit total mass 
value for Cl1449 (see Section~\ref{chewing}). We take the Fourier transform of 
this model, according to Eq.~\ref{eq:vis} and add noise to the model complex visibilities 
based on the weights of the observed ones, assuming natural weighting, and merge the ACA 
and ALMA simulated observations. We then perform a point-source extraction with free 
position, as described in Section~\ref{cooking}, which we compare to the observed offsets. 
In the case of the X-ray centroid, we consider both the combined astrometric uncertainty 
of ALMA/ACA and \emph{Chandra} ($\sim$1'') and a more realistic 5'' precision appropriate 
for the extended emission \citep{Val16}. As shown in Fig.~\ref{fig:offset}, we find 
different probabilities for the SZ-BCG and SZ-X-ray offsets, with the latter falling 
within the top 1.5-5\% of realisations.

\section{\label{appendix:carma}31\,GHz observations}

Cl1449 was observed at 31\,GHz with CARMA between March and April 2012, using the 
3.5\,m sub-array in the SL configuration (project c0865; PI~Riechers). This consisted 
of 6 antennas in a $<$20\,m close-packed configuration probing $\sim$2' scales, and 2 
outrigger antennas to add baselines of $>$50\,m and provide $\sim$0.3' resolution for 
point-source subtraction, for a total baseline range of $\sim$4--83\,m. 
The observation covered 11 tracks, resulting in a combined on-source observing time 
of 31.4\,h. Bandpass calibration was performed during each track using the quasars 
J1512--090 and 3C\,279, and complex gain calibration using the radio quasar J1504+104. 
The planet Mars was used as the primary flux calibrator. The data were then reduced 
using the \texttt{Miriad} \citep{Sau95} software package. Imaging with natural baseline 
weighting results in a synthesised beam size of 135$"$$\times$123$"$, while uniform 
baseline weighting provides a 22$"$$\times$16$"$ beam (for comparison, the primary beam 
FWHM of the 3.5\,m antennas is $\sim$11' at 31\,GHz). We find a continuum r.m.s. noise 
limit of 90\,$\mu$Jy/beam across the full 8\,GHz bandwidth.\\

\noindent
Here we fit point sources using the long-baseline data ($uv>50$\,m) from the outrigger 
antennas and subtract them from all visibilities. 
When extrapolating the 92\,GHz $\sim$-190\,$\mu$Jy signal assuming a standard spectral 
shape for the thermal SZ effect, the decrement at 31\,GHz should be $\sim$-32\,$\mu$Jy. 
Consequently we see no detectable signal, as expected given the noise of the data. We 
nevertheless fit the visibilities with the best-fit model to the 92\,GHz data (see 
Section~\ref{coffee}) at fixed positions corresponding to either the cluster's centre 
of mass or the peak of the 92\,GHz decrement. We find $0\pm120$\,$\mu$Jy, which implies 
a 3$\sigma$ ``upper'' limit to the 31\, GHz signal of $\sim$-360\,$\mu$Jy (i.e., about 
ten times the expectation value).

\section{\label{appendix:lensedib}Lensed background}

\begin{figure}[!h]
\centering
\includegraphics[width=0.49\textwidth]{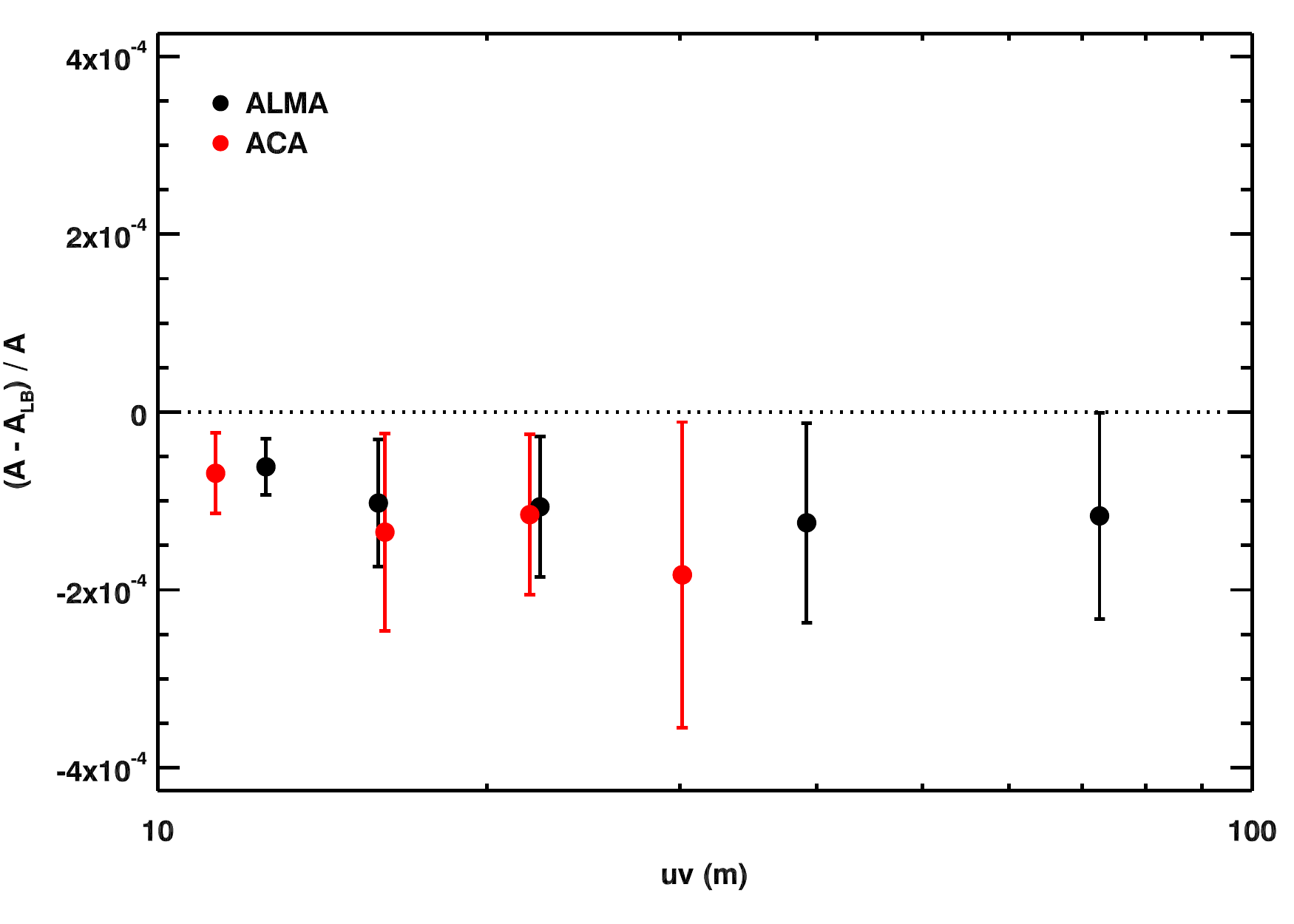}
\caption{Relative amplitude difference, as a function of \emph{uv}-distance, 
between simulated observations of models with (``A$_{\text{LB}}$'') and without (``A'') 
the additional decrement due to lensing-induced number count depletion of undetected 3\,mm 
emitters. As in Fig.~\ref{fig:fft}, the error bars show the r.m.s. deviations between model 
visibilities.}
\label{fig:lensedib}
\end{figure}

Gravitational lensing of background sources by the halo of a galaxy cluster can affect 
its observed SZ decrement in at least two ways: boosting of their flux, which 
contributes to the filling of the decrement, and number count depletion by reducing 
their surface density, which can add signal to the decrement. Here we assume that 
background sources boosted above the detection limit will be identified and subtracted, 
and therefore concentrate on the second effect. 
In this case, the surface brightness of undetected sources (i.e., the background) 
$\Sigma_{\text{IB}}$~can be written as a function of the observed-luminosity function 
$N(S,z)$~and detection threshold $S_{\text{det}}$:\\

\begin{equation}\label{eq:lib}
\Sigma_{\text{IB}}(\theta)\mathrm{d}z = \mu(\theta)^{-1} \mathrm{d}z
\int^{S_{\text{det}}/\mu(\theta)} N(S,z>z_{\text{cluster}})S\mathrm{d}S\text{,}
\end{equation}

\noindent
where $\mu(\theta)$~is the magnification at angular distance $\theta$~from the halo's 
centre \citep[see, e.g.,][]{Brd95,Wri00,Ume14}, assuming spherical symmetry. 
As $\mu(\theta)$ increases with decreasing $\theta$, the infrared background also 
decreases towards the cluster's centre with respect to the unlensed ($\mu=1$) case at 
large radii \citep[e.g.,][]{Zem13,Say18}.
Here we use the 3\,mm number counts distribution of \citet{Zav18}, extrapolating it to 
arbitrarily low fluxes, the redshift distribution of sub-millimetre galaxies of 
\citet[we assume that the number counts distribution is independent of redshift]{Wei13}, 
and a detection limit of five times our ACA r.m.s. Even under this latter conservative 
assumption (the detection limits of both C18 and our ALMA data being lower) we find that 
the contribution of the integrated lensed 3\,mm background to the total decrement is 
negligible compared to the SZ one, as shown in Fig.~\ref{fig:lensedib}.

\section{\label{appendix:fft}\emph{uv} plane modelling}

\begin{figure}[!h]
\centering
\includegraphics[width=0.49\textwidth]{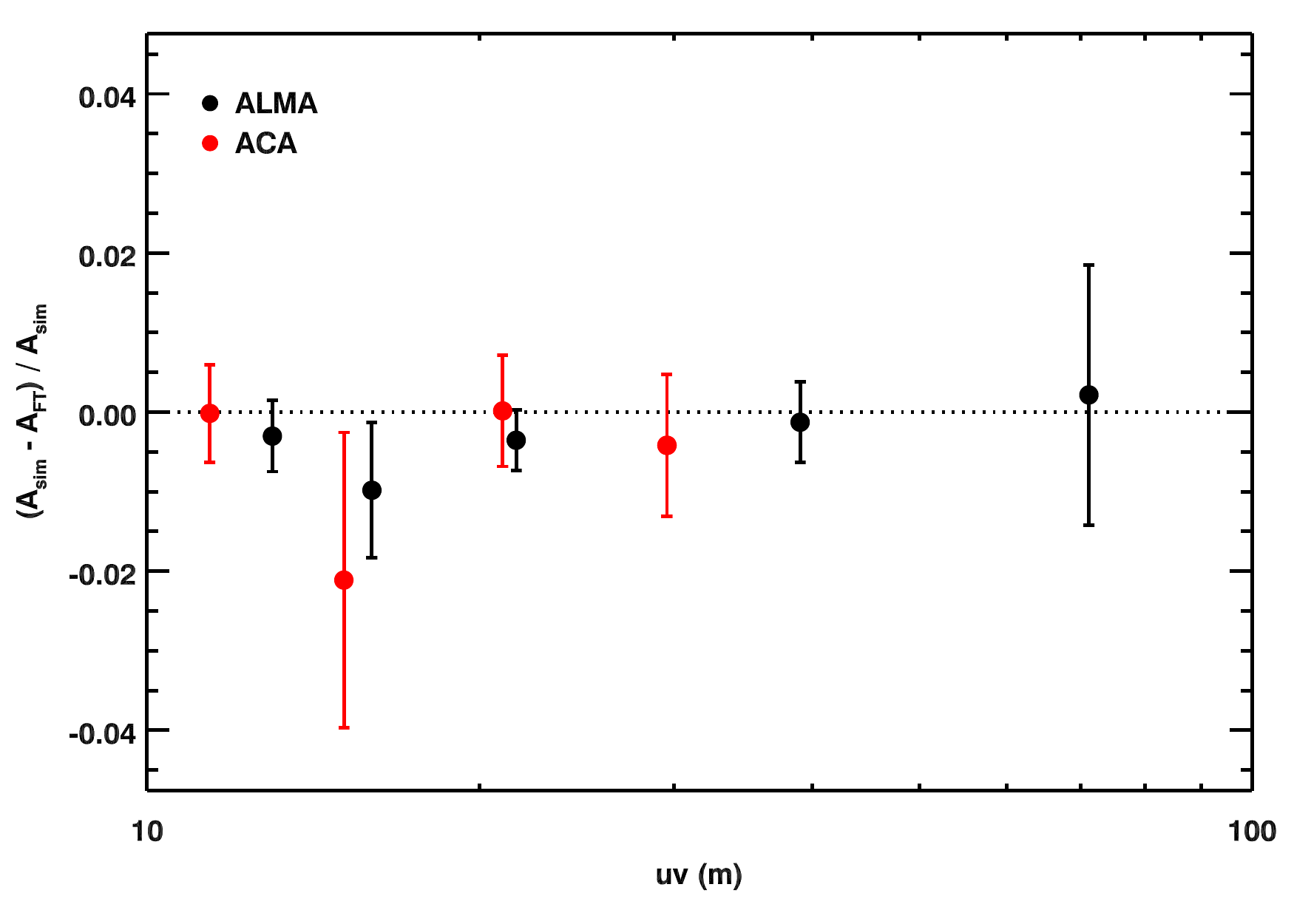}
\caption{
Relative amplitude difference, as a function of \emph{uv}-distance, between the 
output of noise-free simulated observations of parametric models (see main text) 
done with the \texttt{simalma} task of the CASA software and simpler Fourier 
transforms of the same models (with primary beam attenuation) interpolated at the 
same (\emph{u,v}) coordinates. The error bars show the r.m.s. deviations between 
model visibilities computed with both methods.
}
\label{fig:fft}
\end{figure}

For single-pointing observations, the complex visibilities can be approximated as\\

\begin{equation}\label{eq:vis}
V(u,v) = \iint B(x,y)M(x,y)e^{2\pi i(ux+vx)}\mathrm{d}x\mathrm{d}y\text{,}
\end{equation}

\noindent
where $M$~is the on-sky intensity distribution of the model and $B$~the primary 
beam response of the antennae. We compare this to the output of noise-free (\texttt{pwv=0} 
option in \texttt{simalma}) simulated \texttt{simalma} observations, using the GNFW models 
described in Section~\ref{chewing}. 
We find a relative difference between both of at most $\sim$4\% (Fig.~\ref{fig:fft}), 
which is well below the noise level of our data. The approximation given in 
Equation~\ref{eq:vis} therefore allows us to quickly explore the parameter space of models 
at little to no cost of precision, given our S/N (see Fig.~\ref{fig:modfit}). On the other 
hand, while the full observation model used by \texttt{simalma} is certainly more accurate, 
a single iteration requires significantly more time and thus makes automation less 
feasible.\\

\section{\label{appendix:beta}$\beta$-model fit}

\begin{figure}[!h]
\centering
\includegraphics[width=0.49\textwidth]{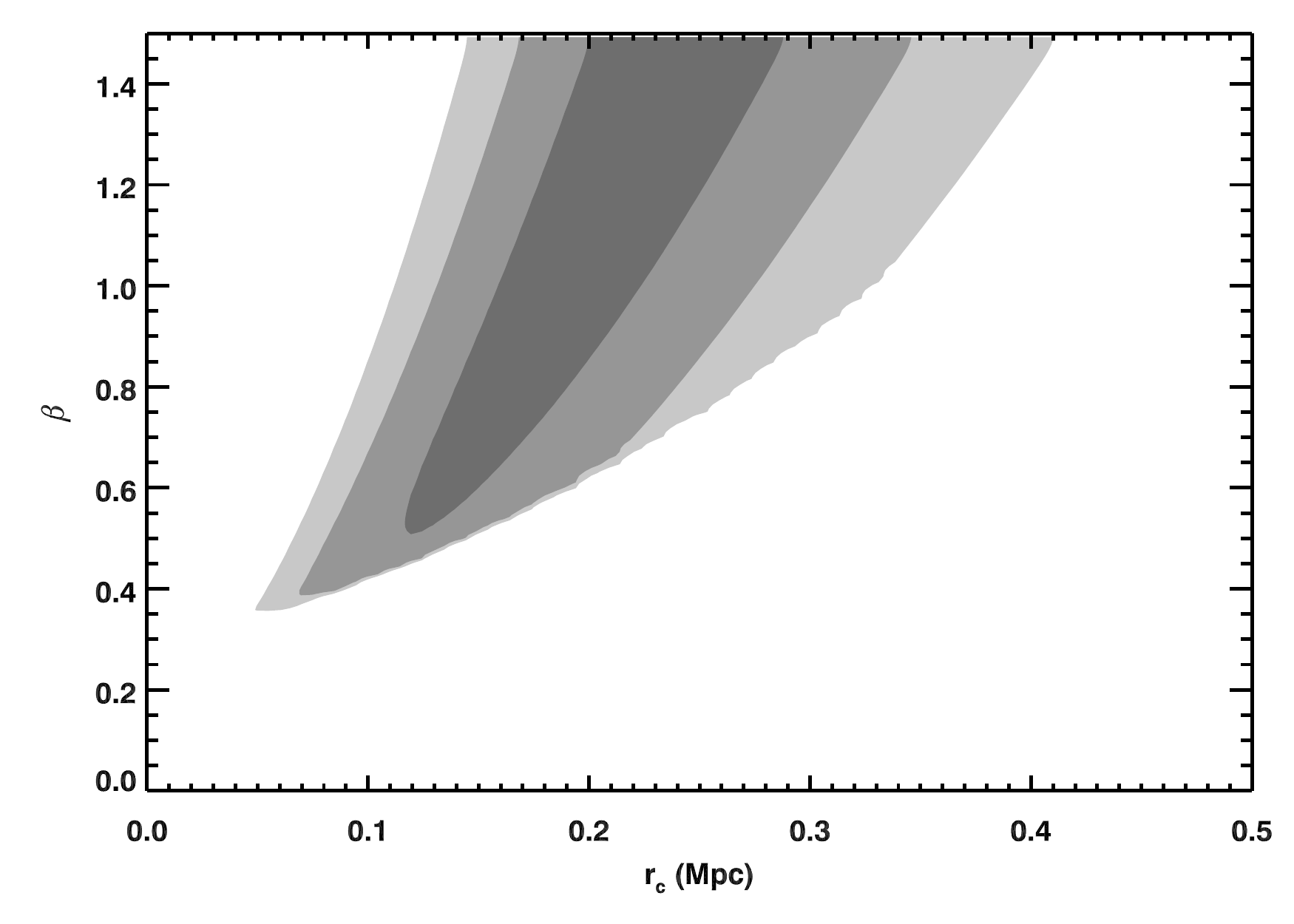}
\caption{$\chi^2$~confidence intervals for the parameters (core radius $r_c$ and index 
$\beta$) of the $\beta$-model fit to the \emph{uv}-amplitude profile. The dark to light 
shaded regions show, respectively, the $1-, 2-, \text{and}~3\sigma$~confidence intervals.
}
\label{fig:betacontours}
\end{figure}

To explore a larger range of pressure profiles we substitute, in the method described 
in Section~\ref{chewing}, the GNFW profiles with a deprojected $\beta$-model for the 
gas density:\\

\begin{equation}\label{eq:beta}
\centering
P(r) = P_0\left[1+\left(\frac{r}{r_c}\right)^2\right]^{-3\beta/2}\text{,}
\end{equation}

\noindent
where $r_c$~is the core radius and $\beta$~the outer slope of the model. We then assume 
primordial abundances and assume the gas fraction and temperature model A from \citet{DR15}. 
We use an expanding parameter grid with $(r_c>0.1,\beta>0.1)$~and steps of 
$(\Delta r_c,\Delta\beta)=(0.01,0.01)$. The result of the fit is shown in 
Fig.~\ref{fig:betacontours}.

\end{appendix}

\end{document}